\documentclass[aps,twocolumn,amsmath,amssymb, superscriptaddress]{revtex4}
\usepackage[pdftex]{graphicx}
 \usepackage{amsmath}
\usepackage[T1]{fontenc}
\usepackage{amssymb}
\usepackage{amsfonts}
\usepackage{color}
\usepackage{bm}
\usepackage[capitalize]{cleveref}
 \usepackage{amsmath} 
\usepackage{lipsum}
\usepackage{amsfonts} 
\usepackage{amssymb, mathrsfs}
\usepackage{braket}
\usepackage{graphicx} 
\usepackage{subfigure}
\usepackage{bbm}
\def\beq{\begin{equation}}
\def\eeq{\end{equation}}
\def\bsp{\begin{split}}
\def\esp{\end{split}}
\def\bea{\begin{eqnarray}}
\def\eea{\end{eqnarray}}
\def\ba{\begin{array}}
\def\ea{\end{array}}

\def\dg{\dagger}

\def\lb{\left(}
\def\rb{\right)}

\def\l.{\left.}
\def\r.{\right.}

\def\ra{\rangle}
\def\la{\langle}

\def\bo{\bold{k}}

\begin{document}

\date{\today}
\title{Weyl magnons in noncoplanar stacked kagom\'e antiferromagnets }
\author{S. A. Owerre}
\affiliation{Perimeter Institute for Theoretical Physics, 31 Caroline St. N., Waterloo, Ontario N2L 2Y5, Canada.}


\begin{abstract}
Weyl nodes have been experimentally realized in photonic, electronic, and phononic crystals. However, magnonic Weyl nodes are yet to be seen experimentally.  In this paper, we propose  Weyl magnon nodes in noncoplanar stacked frustrated kagome antiferromagnets, naturally available in various real materials.  Most crucially, the Weyl nodes in the current system occur at the lowest excitation and possess a topological  thermal Hall effect, therefore they are experimentally accessible at low temperatures due to the population effect of bosonic quasiparticles. In stark contrast to other magnetic  systems, the current Weyl nodes do not rely on time-reversal symmetry breaking by the magnetic order. Rather, they result from explicit macroscopically broken time reversal symmetry by the scalar spin chirality of noncoplanar spin textures, and can be generalized to chiral spin liquid states. Moreover, the scalar spin chirality gives a real space Berry curvature which is not available in previously studied magnetic Weyl systems.   We show the existence of magnon arc surface states connecting projected Weyl magnon nodes on the surface Brillouin zone. We also uncover  the first realization of triply-degenerate nodal magnon point in the non-collinear regime with zero scalar spin chirality.
\end{abstract}
\maketitle

\section{Introduction}
Geometrically frustrated kagom\'e antiferromagnets  are the most studied quantum magnetic systems in condensed matter physics, due to their unconventional properties such as the possibility of quantum spin liquids \cite{nor, zho, sav}, where    magnetic long-range  order  is  forbidden by frustrated interactions down to the lowest temperatures. However, emerging experimental studies have shown that various frustrated kagom\'e antiferromagnets show evidence of intrinsic or magnetic-field-induced magnetic long-range  order  at low temperatures \cite{sup1a, han, han1, zhe,zhe1}. The presence of magnetic long-range  order in frustrated kagom\'e antiferromagnets can be  as a result of  intrinsic perturbative anisotropy such as the Dzyaloshinskii-Moriya interaction  (DMI) \cite{dm,dm2}. The DMI is a consequence of spin-orbit coupling (SOC) and it is present in magnetic systems that lack an inversion center. Interestingly, it is intrinsic to kagom\'e materials \cite{men1,men3, sup1a, han, han1, zhe,zhe1, zor}.  In the magnetically ordered phase, magnons are the quasiparticle excitations. They are uncharged quasiparticles and obey Bose statistics. Most importantly, they  can transport heat, spin current, and carry an intrinsic spin of $1$. These properties make magnons potential candidates for  spintronics and magnetic data storage applications \cite{magn}.

Recently, the concepts of electronic Weyl semimetals (WSMs) \cite{wan,bur} have emerged as the new theme in condensed matter physics, which have been realized experimentally \cite{xu,lv},  and possess numerous potential practical applications.  They are the first realization of Weyl fermions in nature. In principle, Weyl nodes  are allowed in three-dimensional (3D) solid-state crystals with either broken  inversion symmetry or time-reversal symmetry (TRS). Nonetheless,  materials hosting intrinsic Weyl nodes are elusive in nature. In fact, the first experimental evidence of Weyl nodes was realized in an artificial photonic  crystal \cite{lu}, and recently in phononic  crystal \cite{fee}, both of which have the same Bose statistics as magnons.  Recently, this concept has been theoretically extended to magnon bands in the breathing pyrochlore antiferromagnets  \cite{fei} and collinear ferromagnets \cite{mok,su,su1,su2}. Although  bosonic Weyl nodes must occur at finite energy, the Weyl nodes in the lowest excitations have been proven to be the most realistic (see refs. \cite{lu,fee}). However, the predicted  Weyl magnon (WM) nodes in pyrochlore (anti)ferromagnets involve nodes  above the lowest excitation and have not been seen experimentally \cite{mena,mena1}. Therefore, experimentally feasible Weyl magnon nodes are indeed desirable. 

 Thus far,  the breathing pyrochlore antiferromagnet  is the only antiferromagnetic system  exhibiting WM nodes \cite{fei}, but the existence of WM nodes in this system is not very clear as they rely on broken TRS by the magnetic order.  As every magnetically ordered  system breaks TRS, we definitely do not expect all of them to exhibit Weyl nodes.  Moreover,  recent study also shows that the DMI can induce gapped   topological magnon bands in pyrochlore antiferromagnets \cite{lau}.  Hence, it is valid to say that the WMs in pyrochlore antiferromagnet may not be robust against the DMI \cite{foot}, which contradicts the concept of Weyl nodes as robust topological objects. Therefore the mechanism for WMs to exist in antiferromagnetic systems is still an open question and requires further investigation.

In this paper, we predict the existence of different, robust WMs in stacked  frustrated kagom\'e antiferromagnets naturally available in real materials. In contrast to other systems, the current system is  endowed with a  $120^\circ$ non-collinear spin structure  due to  intrinsic out-of-plane DMI or easy-plane anisotropy or intralayer antiferromagnetic  next-nearest-neighbour (NNN) interaction, as realized in real frustrated kagom\'e materials \cite{sup1a,han,han1,zhe,zhe1}. As the conventional  $120^\circ$ non-collinear spin structure   has zero scalar spin chirality and  preserves certain symmetries of the kagom\'e lattice, we find no WMs despite broken TRS by the magnetic order. This is in contrast to previous speculation in pyrochlore antiferromagnets \cite{fei}. However, the magnon bands in this case form doubly-degenerate nodal-line magnons (DDNLMs) and triply-degenerate nodal magnon points (TDNMPs). We argue that the nodal-line magnons  are due to an ``effective TRS'' (i.e.~time-reversal symmetry plus spin rotational and/or mirror reflection symmetry), which is preserved by the  conventional $120^\circ$ non-collinear spin structure. The new TDNMPs or three-component bosons have not been previously proposed in insulating magnets, but they have been realized in fermionic systems  \cite{brad,weng, weng1,lv1}, as condensed matter quasiparticles are not constrained by Lorentz invariance. In stark contrast to DDNLMs in ferromagnets with zero DMI \cite{su1,mok1}, the new TDNMPs  in the conventional $120^\circ$ non-collinear spin structure require the out-of-plane DMI for stability.

The existence of  Weyl nodes  do not necessarily require any special symmetry protection.   Weyl nodes appear in pairs of  opposite chirality, and  can be separated in momentum space when TRS is broken \cite{bur}. In the current study, a slightly canted $120^\circ$ spin structure along the out-of-plane stacking direction form  noncoplanar chiral spin texture with a nonzero scalar spin chirality, which  breaks TRS explicitly and macroscopically. In real materials,  noncoplanar chiral spin texture can be induced either by an external magnetic field applied normal to the conventional $120^\circ$ non-collinear spin structure or by a small in-plane DMI due to lack of mirror symmetry.  In both cases the noncoplanar spin canting has the same effect. Therefore, it suffices to consider only the former case. This leads to the decay of nodal-line magnons and triple magnon points into pairs of WM nodes, which form Weyl cones on the $(010)$ surface Brillouin zone (BZ). They come in pairs of  opposite chirality  and possess chiral magnon surface states (magnon arc) connecting two projected WM nodes on the $(010)$ surface. 

 Furthermore, in contrast to pyrochlore systems \cite{fei,mok,su}, the current WM nodes come from the lowest excitation, which makes them  experimentally feasible due to the population effect of bosonic quasiparticles at low temperatures, and they carry the dominant contribution to the topological  thermal Hall effect. They are, indeed,   robust   as all the dominant intrinsic perturbations have been taken into account. The current results show that robust WMs require macroscopically broken TRS as opposed to broken TRS by the magnetic order in pyrochlore antiferromagnets.  We also find that the WM nodes are always present in the noncoplanar regime provided the interlayer coupling is nonzero. This establishes that both 3D and quasi-2D frustrated kagom\'e antiferromagnets are candidates for investigating antiferromagnetic WMs. As the interlayer coupling always exist in realistic  kagom\'e antiferromagnetic  materials, the current prediction of WMs can be experimentally accessible by inelastic neutron scattering experiments. We note that topological antiferromagnets have potential technological applications in spintronics \cite{pea}. They  have zero spin magnetization which makes their magnetism  externally invisible and insensitive to external magnetic fields, therefore are more efficient in spintronics applications.

 \section{Spin Model}
\begin{figure}
\centering
\includegraphics[width=1\linewidth]{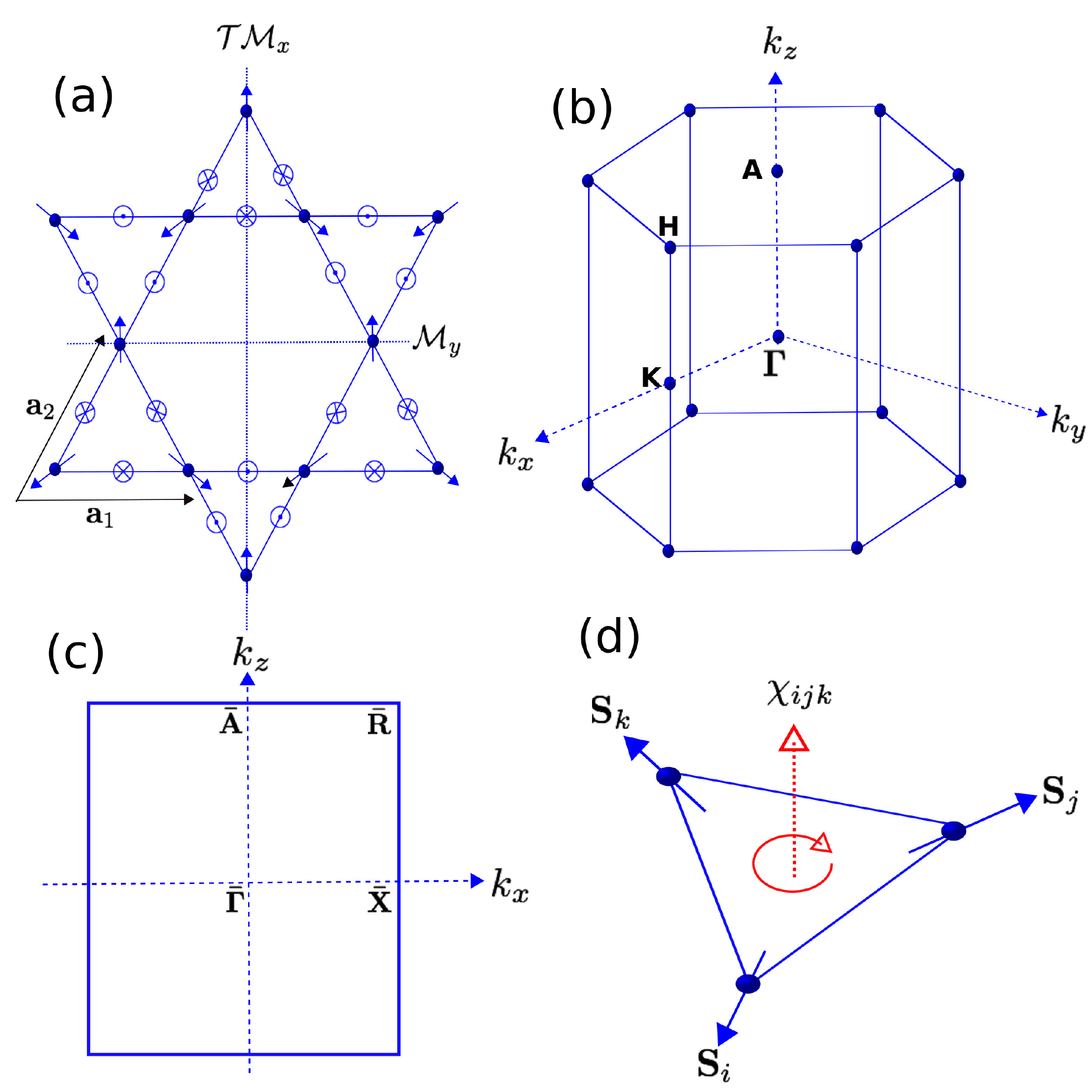}
\caption{ {(a)} Top view of  unshifted kagom\'e lattice stacked along the (001) direction.  The $120^\circ$ non-collinear spin configuration with a positive vector chirality is indicated (blue arrows). The in-plane unit vectors are ${\bf a}_1=(1,0,0)$ and ${\bf a}_2=(1/2,\sqrt{3}/2,0)$. The unit vector along the stacking direction ${\bf a}_3=(0,0,1)$ is not depicted. The mirror reflection axes (dotted lines) and the direction of DMI (crossed and dotted circles) are indicated.  { (b)} Bulk Brillouin zone (BZ) of stacked hexagonal lattice with indicated high-symmetry points.  {(c)} (010) surface BZ. { (d)} Scalar spin chirality of three noncoplanar spins on a triangle:  $\chi_{ijk}=  {\bf S}_{i}\cdot \lb {\bf S}_{j}\times{\bf S}_{k}\rb$ (dash red arrow).}
\label{KL}
\end{figure}

  We study  stacked  frustrated kagom\'e antiferromagnets. Generally, they are governed by the microscopic spin Hamiltonian 
\begin{align}
\mathcal H&=J\sum_{\la ij\ra,\ell} {\bf S}_{i,\ell}\cdot{\bf S}_{j,\ell}+\sum_{\la ij\ra,\ell}{\bf D}_{ij}\cdot {\bf S}_{i,\ell}\times{\bf S}_{j,\ell}\nonumber\\&+ J_c\sum_{i,\la \ell \ell^\prime\ra}{\bf S}_{i,\ell}\cdot{\bf S}_{i,\ell^\prime}- H\sum_{i,\ell} S_{i,\ell}^z,
\label{ham}
\end{align}
where $i$ and $j$ denote the sites on  the kagom\'e layers, $\ell$ and $\ell^\prime$ label the layers.  The first term is an intralayer antiferromagnetic nearest-neighbour (NN) Heisenberg interaction denoted by $J$.  The second term is the DMI, where ${\bf D}_{ij}$ is the DM vector between site $i$ and $j$,  due to lack of inversion symmetry between two sites on each kagom\'e  layer. It is a perturbative anisotropy to the Heisenberg interaction $J$, and it is dominated by the out-of-plane component  (i.e. ${\bf D}_{ij}=\pm D_z{\bf \hat z}$). The DMI alternates between the triangular plaquettes of the kagom\'e lattice as shown in Fig.~\eqref{KL}a. The out-of-plane DMI stabilizes  the $120^\circ$ non-collinear spin structure and its sign  determines the vector chirality of the non-collinear spin order \cite{men1,men3}.  The third term  is an interlayer NN  Heisenberg interaction  between the kagom\'e  layers denoted by $J_c$. It can either be  ferromagnetic ($J_c<0$) or antiferromagnetic ($J_c>0$). The last term  is an external magnetic field with strength $H$ in units of $g\mu_B$, and it is applied along the stacking direction taken as the $z$-axis. Without loss of generality, we consider unshifted  stacked kagom\'e layers as  realized in different stacked frustrated kagom\'e antiferromagnets \cite{han,han1,zhe,zhe1}. Note that  all the interactions in Eq.~\eqref{ham} are present in real physical  kagom\'e  materials, and the magnetic field is readily available in the laboratories. In the following, we consider antiferromagnetic  interlayer coupling ($J_c>0$). The   ferromagnetic interlayer coupling ($J_c<0$) will be briefly discussed  in the Appendixes.

\begin{figure*}
\centering
\includegraphics[width=1\linewidth]{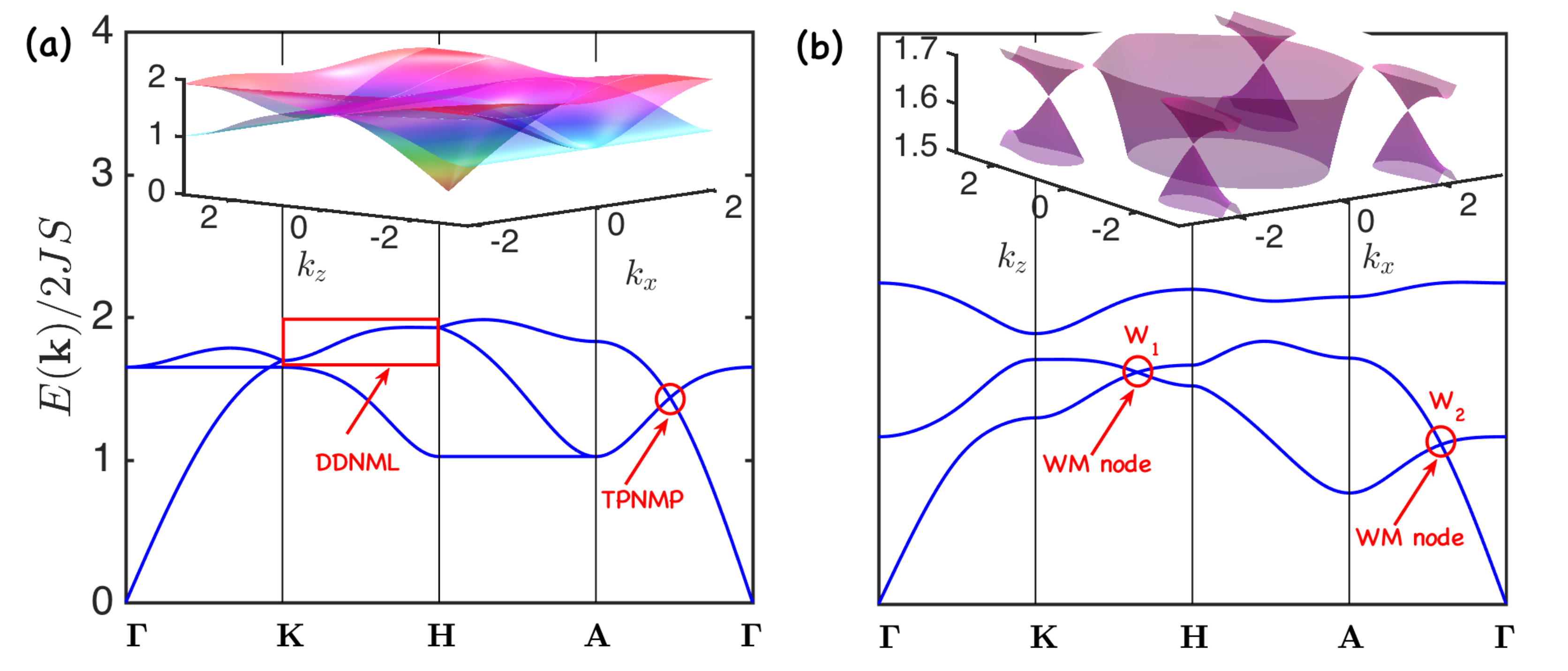}
\caption{ Magnon bands of stacked kagom\'e antiferromagnets. (a) Magnon band structure of stacked 120$^\circ$ non-collinear spin structure (with zero scalar spin chirality) showing doubly-degenerate nodal-line magnons (red rectangle) and triply-degenerate nodal magnon point (red circle) for  $D_z/J = 0.2$, $J_c/J = 0.5$, $H =0$.  The flat bulk magnon band along ${\bf H}$--${\bf A}$ line is a lifted zero energy mode due to the presence of the DMI. It is an artifact of the kagom\'e-lattice structure and can acquire a small dispersion upon the inclusion of an intralyer antiferromagnetic NNN interaction. Inset shows 3D magnon band in the $k_y=0$ plane with nodal magnon rings in the $k_x$-$k_z$ momentum space. (b) Magnon band structure of stacked noncoplanar spin structure (with nonzero scalar spin chirality) showing Weyl magnon nodes   for $D_z/J = 0.2$, $J_c/J = 0.5$, $H =0.3H_s$, where $H_s= 6J +2\sqrt{3}D_z+4J_c$ is the saturation field.  Note that the  flat bulk magnon band is now dispersive in the noncoplanar regime, and the doubly-degenerate nodal-line magnons  and the triply-degenerate nodal magnon point are lifted with the appearance of WM nodes. Inset shows 3D magnon band in the $k_y=0$ plane for W$_1$  Weyl cones. }
\label{band}
\end{figure*}

\begin{figure}
\centering
\includegraphics[width=1\linewidth]{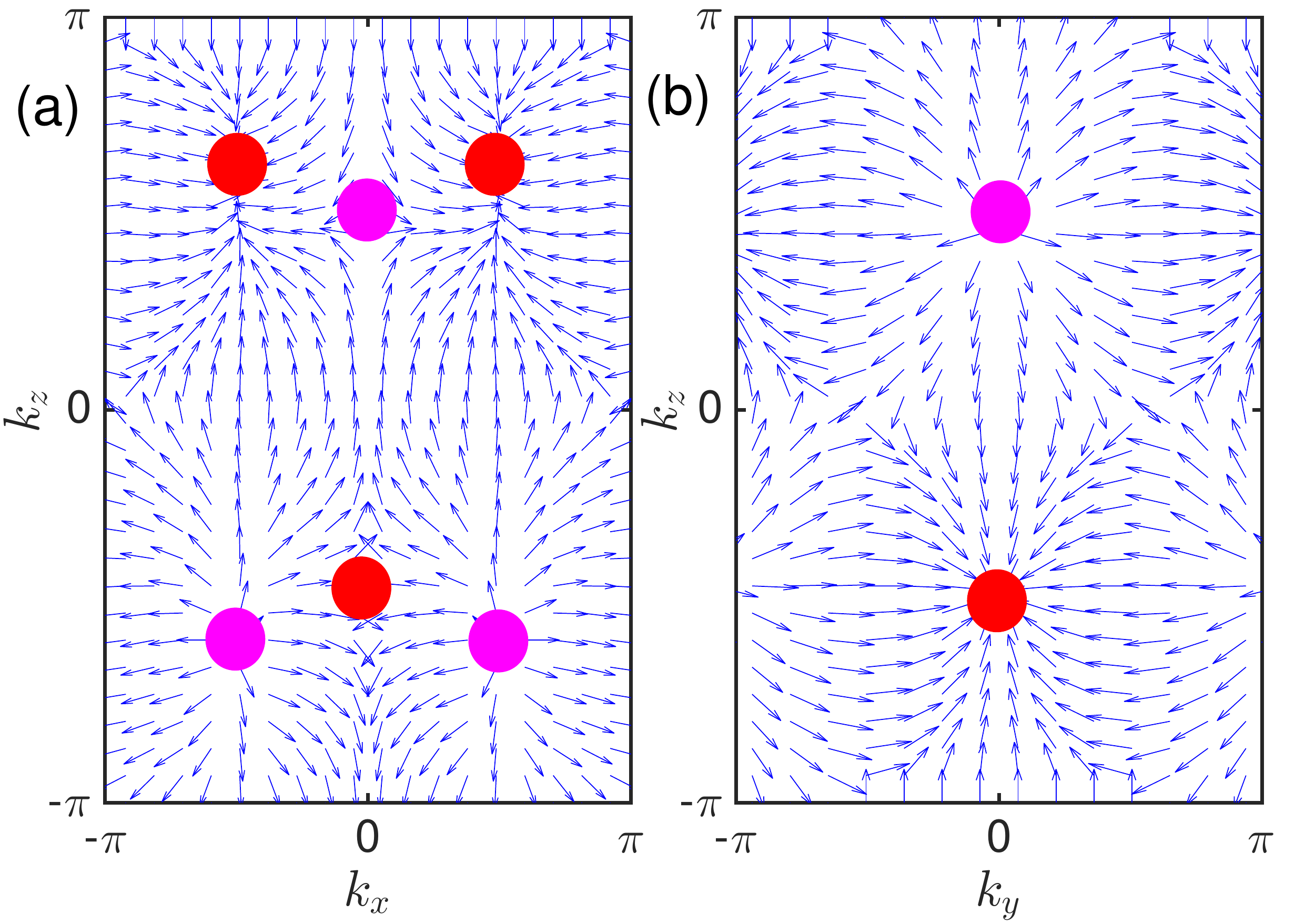}
\caption{ Monopole distribution of the lowest magnon band Berry curvature $\Omega_{1,xz}^{y}(\bo)$ on the $k_y=0$ plane (a) and $\Omega_{1,yz}^{x}(\bo)$ on the $k_x=2\pi/3$ plane (b), showing the monopole (red dot) and anti-monopole (pink dot) distribution of WM nodes. The parameters are the same as Fig.~\ref{band}(b).}
\label{BC}
\end{figure}


\section {Nodal-line magnons in the non-collinear spin structure}
We commence our study in the limit of zero magnetic field, $H=0$. In this limit, the classical ground state of the Hamiltonian Eq.~\eqref{ham} is a stacked $120^\circ$ non-collinear spin  structure with positive vector chirality and zero scalar spin chirality. Now, we study the magnon excitations  of this spin structure. In order to do this, we rotate our spin quantization axis locally in spin space so that it aligns with the magnetic ordering, and then we introduce the  Holstein-Primakoff bosons (see Appendixes~\eqref{appena} and \eqref{appenb}).  If the DMI is also set to zero  (i.e. $D_z=0$), the magnon bands of stacked $120^\circ$ non-collinear spin  structure show zero energy modes for a constant  out-of-plane momentum $k_z$ along ${\bf H}$--${\bf A}$ line \cite{sch,sch1}.  Therefore, the absence of magnetic long-range  order in the 2D frustrated kagom\'e antiferromagnets also persists in the 3D limit. As shown in Fig.~\ref{band}(a),  a nonzero DMI (i.e. $D_z\neq 0$) lifts the zero energy mode to a flat magnon band along ${\bf H}$--${\bf A}$ line, and thus stabilizes the stacked (3D)  $120^\circ$ non-collinear spin configuration as in the 2D system \cite{men1}.

In addition, the magnon bands of stacked $120^\circ$ non-collinear spin  structure (i.e. $J_c\neq 0$) form DDNLMs along ${\bf K}$--${\bf H}$ and TDNMPs  along ${\bf A}$--${\bf \Gamma}$ lines of the BZ (Fig.~\ref{KL}(b) and (c)) as depicted in Fig.~\ref{band}(a).  In the weakly coupled realistic regime $J_c/J<1$, there are DDNLMs along ${\bf K}$--${\bf H}$ and TDNMPs along ${\bf A}$--${\bf \Gamma}$. However, in the strongly coupled regime  $J_c/J>1$ (probably unrealistic), there are only TDNMPs along ${\bf K}$--${\bf H}$ and ${\bf A}$--${\bf \Gamma}$ lines (not shown). The  TDNMPs or three-component bosons are the analogs of fermionic counterparts  \cite{brad,weng,weng1,lv1}. They are  allowed as condensed matter quasiparticles are not constrained by Lorentz invariance.

 The  DDNLMs and TDNMPs in the conventional $120^\circ$ non-collinear spin structure can be understood as follows.  For a perfect kagom\'e lattice with strong out-of-plane DMI the ground state of the Hamiltonian Eq.~\eqref{ham} at zero field is a $120^\circ$ non-collinear spin structure with positive vector chirality and zero scalar spin chirality as depicted in Fig.~\eqref{KL}a. This non-collinear spin structure preserves all the symmetries of the kagom\'e lattice. For instance, the combination of TRS (denoted by $\mathcal T$)  and spin rotation denoted by $\mathcal R_z(180^\circ)$ is a good symmetry. Here, $\mathcal R_z(180^\circ)=\text{diag}(-1,-1,1)$ denotes  a $180^\circ$ spin rotation of the in-plane coplanar spins about the $z$-axis, and `\text{diag}' denotes diagonal elements.  The system also has three-fold rotation symmetry along the $z$ direction denoted by  $\mathcal C_{3}$. Moreover, mirror reflection symmetry of the kagom\'e plane about the $x$ or $y$ axis in combination with $\mathcal T$ (i.e. $\mathcal T\mathcal M_x \mathcal T $ or $\mathcal M_y \mathcal T $ ) is also a symmetry of the $120^\circ$ non-collinear spin structure. These symmetries are referred to as an ``effective TRS'' and lead to  DDNLMs and TDNMPs.  They are different from the DDNLMs in insulating ferromagnets \cite{mok1,su1}  by the presence of the DMI and TDNMPs.
  \begin{figure*}
\centering
\includegraphics[width=1\linewidth]{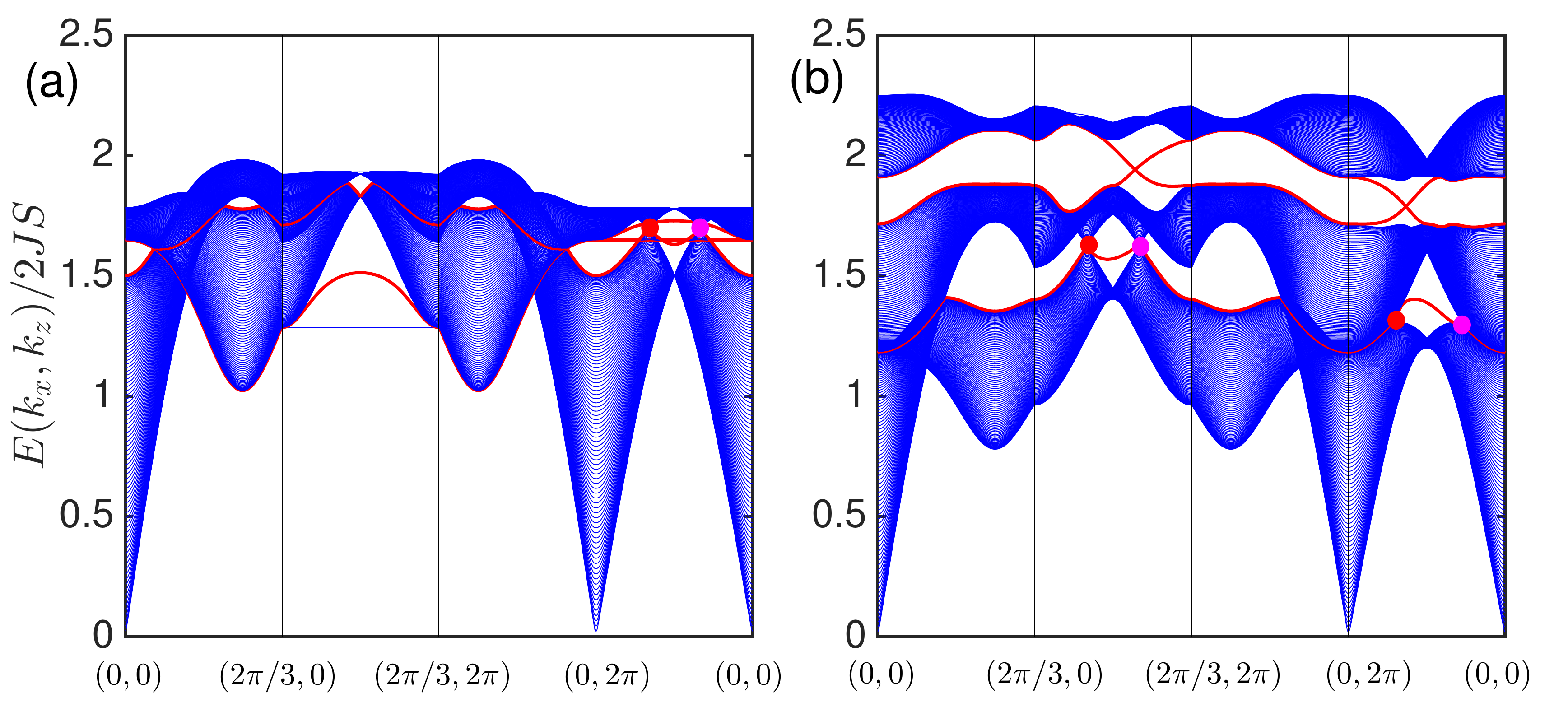}
\caption{  { (a)} (010)-projected  nodal-line magnons (red and pink dots)  with drumhead magnon surface states  (red lines) at zero scalar spin chirality with the parameters of Fig~.\ref{band}(a). The (010)-projected flat bulk magnon band in the noncollinear regime (i.e. zero scalar spin chirality) is a lifted zero energy mode due to the presence of the DMI. It is an artifact of the kagom\'e-lattice structure and can acquire a small dispersion upon the inclusion of an intralyer antiferromagnetic NNN interaction}.    {(b)}. (010)-projected  Weyl magnon nodes  of opposite chirality (red and pink dots)   connected by  chiral magnon arc surface states  (red lines) at nonzero scalar spin chirality with the parameters of Fig~.\ref{band}(b). 
\label{tss}
\end{figure*}
  \begin{figure}
\centering
\includegraphics[width=1\linewidth]{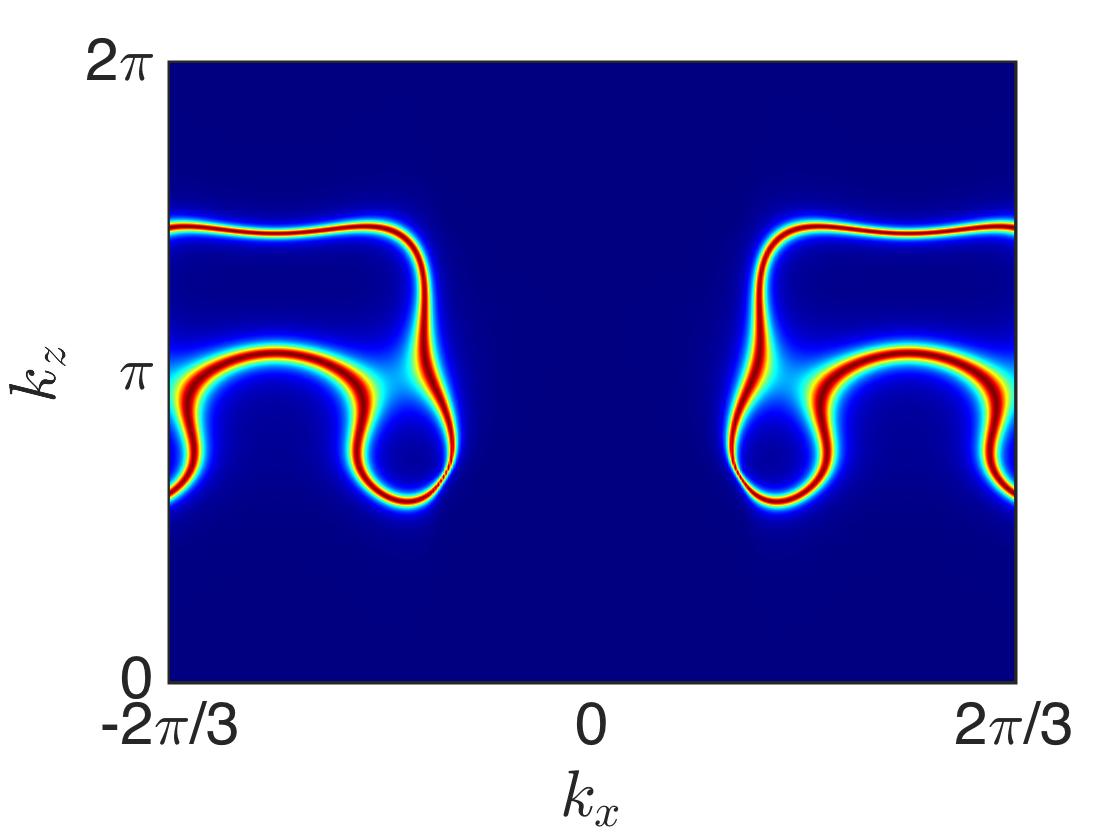}
\caption{ The density plot of the magnon surface spectral function on the (010) surface BZ. The W$_1$ Weyl nodes form magnon arcs at $E\approx E_{W_1}$. The parameters are the same as Fig~.\eqref{band}b.}
\label{arc}
\end{figure}
 
 \section{ Weyl  magnons in the noncoplanar spin structure}
 As  in  electronic systems,  Weyl nodes  do not necessarily require any special symmetry protection. They are formed by non-degenerate bands and appear in pairs of  opposite chirality, and can be separated in momentum space when  TRS is broken. Now, we  will break the  ``effective TRS'' and study its effect on the magnon bands.  There are two ways in which this  symmetry  can be broken.  The first one is intrinsic when the kagom\'e lattice lacks a mirror reflection symmetry. Therefore, a small in-plane DMI will be allowed and  induces   noncoplanar chiral spin textures \cite{sup1a}. However, in most frustrated kagom\'e antiferromagnets, e.g. herbertsmithite \cite{zor}, the in-plane DMI can be very weak and negligible.  The  second one is extrinsic by applying an out-of-plane external magnetic field perpendicular to the in-plane stacked $120^\circ$ non-collinear spin structure. This  also induces noncoplanar chiral spin textures with a nonzero scalar spin chirality given by $\chi_{ijk;l}=  {\bf S}_{i,\ell}\cdot\lb {\bf S}_{j,\ell}\times{\bf S}_{k,\ell}\rb$ as shown in  Fig.~\ref{KL}(d) (see Appendix~\eqref{appena}). 
  
Interestingly,  the magnetic-field-induced  $\chi_{ijk;l}$  can persist even when the $120^\circ$ non-collinear spin structure is stabilized by other perturbative interactions different from the DMI, e.g. easy-plane anisotropy \cite{che} or intralayer antiferromagnetic NNN interaction\cite{har1}. Therefore, we expect that WMs should also exist in stacked noncoplanar chiral antiferromagnets without the DMI.  This noncoplanar chiral spin texture breaks TRS macroscopically, hence we can now look for the existence of WMs. In Fig.~\ref{band}(b) we plot the magnon bands in the noncoplanar regime. Evidently, we can see  that the DDNLMs and TDNMPs give way for WM nodes, which are formed by linear crossing of two non-degenerate magnon bands  at isolated points in momentum space.   We denote the locations of the WM nodes along the $k_z$ momentum direction by $k_{W_i}$. Please see Appendix~\eqref{appenc} for the long expressions for  $k_{W_i}$.

In particular, the lowest and middle non-degenerate magnon bands cross linearly in the weakly-coupled realistic limit $J_c<J$, and form three pairs of WM cones on the (010) surface BZ located at $(\pm 2\pi/3,0,\pm k_{W_1})$  and $(0,0,\pm k_{W_2})$.  In the strongly-coupled regime $J_c\geq J$ (probably unrealistic), we also find that the lowest and middle magnon bands cross linearly at $(\pm 2\pi/3,0,\pm k_{W_1})$  and $(0,0,\pm k_{W_2})$.  In addition, the topmost and lowest magnon bands cross linearly at $(\pm 2\pi/3,0,\pm k_{W_3})$ and $(0,0,\pm k_{W_4})$ (see Appendix~\eqref{appenc}). In this case,  the W$_3$ WM nodes  form  the analog of type-II WSM \cite{soluy} (see Appendix~\eqref{appenc}).  Most importantly, we find that WM nodes always exist in the noncoplanar regime for all nonzero values of $J_c/J$ (see Appendix~\eqref{appenc}). This suggests that both 3D and quasi-2D stacked kagom\'e antiferromagnets are candidates for WM nodes.

 It is crucial to point out that the WM nodes in noncoplanar stacked frustrated kagom\'e antiferromagnets are different  from those of breathing pyrochlore antiferromagnets \cite{fei}. The latter requires no DMI and rely on broken TRS by the magnetic order. However, as every magnetically ordered system breaks TRS, we do not expect every ordered magnetic system to have WM nodes.  Moreover, gapped topological magnon bands were recently found in pyrochlore antiferromagnets with DMI \cite{lau}.  This suggests that the WMs in breathing pyrochlore antiferromagnets may not be robust when all the proper DMIs are taken into account \cite{foot}.  In the current study, however,  the only requirement for the existence of WM nodes  is the macroscopically broken TRS by $\chi_{ijk;l}$, which can be induced by the in-plane DMI  or an external magnetic field as shown here, in addition to the out-of-plane DMI. They  also persist in the absence of DMI as the $120^\circ$ non-collinear spin structure can be stabilized through  other means, such as an easy-plane anisotropy or an intralyer antiferromagnetic NNN interaction.  As we mentioned above, our results show that WMs should not be allowed in every magnetically ordered system that breaks TRS by the magnetic order.   Indeed, the current WM nodes are robust as all the dominant  intrinsic perturbations have been taken into account. It should be noted that although  bosonic Weyl nodes must occur at finite energy, the Weyl nodes at the lowest excitation are the most realistic (see refs. \cite{lu,fee}), and this is the case in the current system.
 
 \subsection{Weyl magnons as monopoles of the Berry curvature}
 
One of the most important features of Weyl nodes is that they are topological objects. This means that a pair of Weyl nodes cannot be removed by small perturbations. They can only be removed by annihilating each other in momentum space. Weyl nodes also  act as the source and sink of the Berry curvature.  In other words, a single Weyl node can be considered as a monopole of the Berry curvature.  In contrast to other magnetic systems, the Berry curvature in the current system is already present in the real space spin configuration due to $\chi_{ijk;l}$ as depicted in Fig.~\ref{KL}(d). The momentum space Berry curvature of a given magnon band $n$ is defined  as
\begin{align}
\Omega_{n, \alpha\beta}^{\gamma}(\bo)=-\sum_{m\neq n}\frac{2\text{Im}\big[ \braket{\mathcal{P}_{\bo n}|\hat v_\alpha|\mathcal{P}_{\bo m}}\braket{\mathcal{P}_{\bo m}|\hat v_\beta|\mathcal{P}_{\bo n}}\big]}{\big[ E_{n}(\bo)-E_{m}(\bo)\big]^2},
\label{chern2}
\end{align}
where $\hat v_\alpha=\partial \mathcal{H}_B(\bo)/\partial k_\alpha$ defines the velocity operators with $\alpha,\beta,\gamma=x,y,z$; $\mathcal{H}_B(\bo)$ is the  magnon Bogoliubov Hamiltonian, whereas $\mathcal{P}_{\bo n}$ are the paraunitary  operators (eigenvectors) that diagonalize  the Hamiltonian, and  $E_{n}(\bo)$ are the eigenvalues or magnon energy bands (see Appendix~\eqref{appenca} for more details). Note that the Berry curvature is a 3-pseudo-vector pointing along  the $\gamma$ directions perpendicular to both the $\alpha$ and $\beta$ directions.  From the denominator of the Berry curvature in Eq.~\eqref{chern2}, it is evident that it diverges at the WM nodes. As can be clearly seen in Fig.~\eqref{BC}, the WM nodes come in pairs of opposite chirality, and act as source (monopole) and sink (anti-monopole) of the Berry curvature, with $\pm 1$ chirality  (red and pink dots respectively).  

We note that for  $-k_{W_i}<|k_z|<k_{W_i}$, our  system is gapped and can be considered as slices of 2D topological magnon Chern insulators \cite{sol, sol1}. In this case there are definite  Chern numbers (defined as the integration of the Berry curvature over the BZ) in the $k_x$-$k_y$ plane for fixed $k_z$. They are estimated as $C_{1,2}(-k_{W_i}<|k_z|<k_{W_i})=\pm\text{sgn}(\sin(\phi))$ for the first two bands and $C_3(-k_{W_i}<|k_z|<k_{W_i})=0$ for the topmost band, where $\phi$ is the angle subtended by three noncoplanar spins in a unit triangle (see Appendix~\eqref{appenb}), and $\sin\phi$ is proportional to  $\chi_{ijk}$ (see Fig.~\ref{KL}(d)).

\subsection { Magnon arc surface  states}
 One of the hallmarks of Weyl nodes is the Fermi arc surface states. In the current model,  they will be referred to as magnon arc surface  states, and they connect projected bulk WM nodes on the surface BZ.   Let us consider the (010) surface and assume that the system is infinite along $x$ and $z$ directions, so both $k_x$ and $k_z$ are good quantum numbers.   As shown  in Fig.~\ref{tss}(a), the (010)-projected nodal-line magnons in the 120$^\circ$ non-collinear spin structure at zero scalar spin chirality have drumhead magnon surface states (red lines). In the noncoplanar regime the nodal-line magnons give way for WM nodes. As shown in Fig.~\ref{tss}(b), the (010)-projected WM nodes of  opposite chirality (red and pink dots) are connected by  magnon arc surface states. Also note that in this weakly coupled realistic regime ($J_c<J$), the middle and topmost bands also feature a topological magnon Chern insulator  with gapless surface states between the bulk gaps.   The magnon arcs formed by topologically protected surface states that connect projected WM nodes are depicted in Fig.~\eqref{arc}.
 
 \section{Topological thermal Hall effect due to Weyl magnons}
 
Topological Hall effect  refers to the generation of a transverse Hall conductivity  as a result of nontrivial noncoplanar chiral spin textures \cite{ele,ele1,ele0, ohg, sur}.  It can also occur in the absence of SOC or DMI due to the scalar spin chirality. In the current study, this will be referred to as topological thermal Hall effect as it applies to charge-neutral quasiparticles such as magnons. 

Moreover, the most important property of WMs in the current noncoplanar chiral spin textures is that they come from the lowest magnon excitation, hence they contribute significantly to the topological thermal Hall conductivity at low temperatures.  We note that  the WMs in pyrochlore (anti)ferromagnets \cite{mok,su,fei} occur above the lowest excitation at high energy. However, in any bosonic system the lowest excitation is thermally populated at low temperatures due to the Bose function, and makes dominant contributions to the  thermal Hall conductivity \cite{alex0,alex2, alex2a, shin1}.   Hence, due to the population effect the WMs in pyrochlore (anti)ferromagnets \cite{mok,su,fei} will not contribute to the thermal Hall effect at low temperatures. Therefore, previously  experimentally reported thermal Hall conductivity in pyrochlore ferromagnets  \cite{alex1,alex1a}, and a subsequent theoretical calculation \cite{alex2a} are definitely not a consequence of  recently proposed WMs in this system \cite{mok,su}. In this regard,  it is valid to say that the most important WM nodes with potential applications are definitely those at the lowest excitation \cite{foot}.  

In this section, we will show that  the topological  thermal Hall effect of WMs in this system depends on the  distribution  and distance between the WM nodes in momentum space, which is a function of the scalar spin chirality of noncoplanar chiral spin textures. We  note that the topological or anomalous thermal Hall effect induced by WMs has not been studied both theoretically and experimentally. We will provide a theoretical description in this section, and hopefully an experimental probe will be done in the future.   The  thermal Hall effect is due to the flow of heat current $J_\alpha^\gamma$ under the influence of a thermal temperature gradient  $\nabla_{\beta}T$. It can be derived from linear response theory \cite{alex0, alex2a,alex2,shin1}. For the 3D model, the total intrinsic anomalous thermal Hall conductivity can be written as $\kappa_{H}=\lb \kappa_{yz}^x + \kappa_{zx}^y +\kappa_{xy}^z \rb/3$, where
the components $\kappa_{\alpha\beta}^\gamma=-J_\alpha^\gamma/\nabla_{\beta}T$  are  given explicitly by 
\begin{align}
\kappa_{\alpha\beta}^\gamma=- T\int_{{BZ}} \frac{\text{d}\bo}{(2\pi)^3}~ \sum_{n=1}^N c_2\lb f_n^B\rb\Omega_{n, \alpha\beta}^{\gamma}(\bo),
\label{thm}
\end{align}
where   $ f_n^B=\lb e^{E_{n}(\bo)/T}-1\rb^{-1}$ is the Bose function  with the Boltzmann constant set to unity,  and $ c_2(x)=(1+x)\lb \ln \frac{1+x}{x}\rb^2-(\ln x)^2-2\text{Li}_2(-x)$, with $\text{Li}_2(x)$ being the  dilogarithm. Evidently, the anomalous thermal Hall conductivity is the integration of the Berry curvature over the BZ, weighed by the $c_2$ function. Due to the Berry curvature, $\kappa_{\alpha\beta}^\gamma$ is also a 3-pseudo-vector $\lb \kappa_{yz}^x, \kappa_{zx}^y, \kappa_{xy}^z\rb$, pointing along  the $\gamma$ directions perpendicular to both the $\alpha$ and $\beta$ directions. 

  \begin{figure}
\includegraphics[width=1\linewidth]{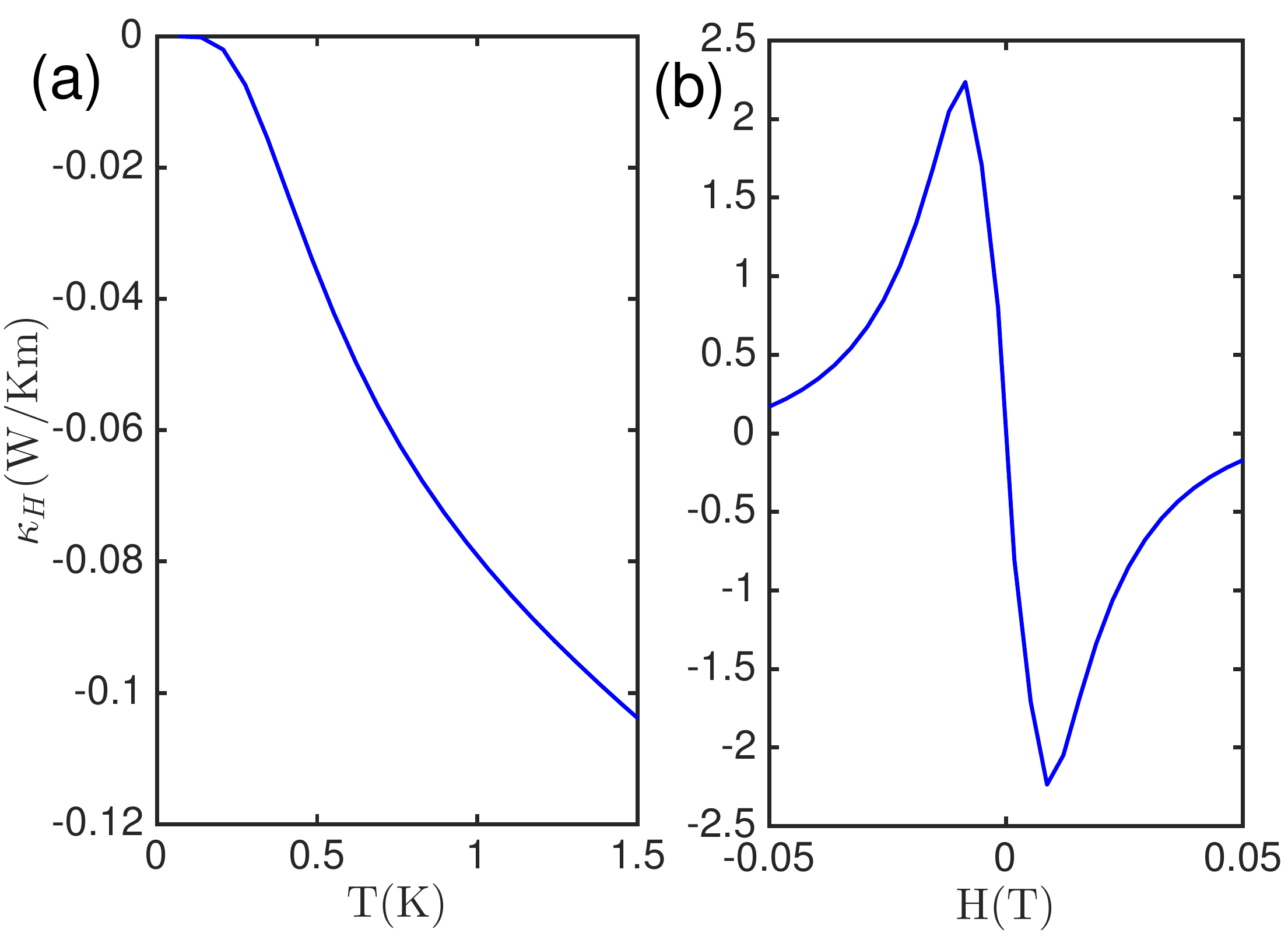}
\caption{Color online.  Topological thermal Hall conductivity. (a) The plot of $\kappa_H$ vs. $T$ for $D_z/J=0.2,~J_c/J=0.5$ and $H=0.3H_s$.  (b) The plot of $\kappa_H$ vs. $H$  for $D_z/J=0.2$, $J_c/J=0.5$, and $T/J=0.75$.}
\label{THE}
\end{figure}

For any surface not perpendicular to the $k_z$ direction, the WM nodes come in pairs of opposite chirality. Therefore, the net Berry curvature vanishes identically.  In  fact, a direct numerical integration of Eq.~\eqref{thm} shows that  $\kappa_{yz}^x = \kappa_{zx}^y\approx 0$. However, for the surface perpendicular to the $k_z$ direction the distributions of WM nodes are different. As we can see from the Berry curvature fields in Fig.~\eqref{BC},  for fixed $k_z$ in the vicinity of any WM nodes the  net Berry curvature in the $k_x$-$k_y$ plane is nonzero. Therefore, integrating over $k_z$ results in nonzero $\kappa_{xy}^z$. Now, we can separate the nonzero component in Eq.~\eqref{thm} as
\begin{align}
\kappa_{xy}^z= \int_{-\pi}^{\pi} \frac{d k_z}{2\pi}\kappa_{xy}^{\text{2D}}(k_z),
\label{ATHC}
\end{align}
  where $\kappa_{xy}^{\text{2D}}(k_z)$ is a set of 2D thermal Hall conductivity in the $k_x$-$k_y$ plane \cite{sol, sol1} parameterized by $k_z$, which is given by
  
  \begin{align}
\kappa_{xy}^{\text{2D}}(k_z)=- T\int \frac{\text{d}\bo_\parallel}{(2\pi)^2}\sum_{n=1}^N c_2\lb f^B[E_n(\bo_\parallel,k_z)]\rb\Omega_{n\bo_\parallel}^z(k_z),
\label{ATHC1}
\end{align}
 where $\bo_\parallel=(k_x,k_y)$. As the Berry curvature is dominant  near the WM nodes, the major contribution  to the thermal Hall conductivity comes from these nodes at the lowest magnon band due to the Bose function. 
 
 Thus,  at  nonzero temperature ($T\neq 0$) the  empirical expression for the topological  thermal Hall conductivity can be written as
 \bea 
 \kappa_{xy}^z\propto \sum \Delta k_0^i,
 \label{sep}
 \eea  
 where $\Delta k_0^i$ is the separation of the WM nodes along the $k_z$ momentum direction, which depends on the scalar spin chirality of noncoplanar chiral spin textures. This  relation is akin to the anomalous Hall conductivity in electronic Weyl semimetal \cite{ya,bur}.  Indeed, when the WM nodes annihilate at the Brillouin zone (BZ) boundary, the system becomes a  fully gapped 3D topological magnon insulator with similar features to 2D counterparts \cite{sol, sol1}. As shown in Fig.~\ref{THE}(a), the topological thermal Hall conductivity vanishes at zero temperature as no magnons are thermal excited. It also vanishes at zero scalar spin chirality  in accordance with Eq.~\eqref{sep}, i.e.  TRS is not broken macroscopically. In Fig.~\ref{THE}(b), we see  that a small magnetic field is capable of changing the sign of the scalar spin chirality, which leads to a sign change in the topological thermal Hall conductivity. 
 
 It is important to note that in magnetic insulators, the thermal Hall effect is a direct consequence of the Berry curvature. This means that the thermal Hall effect is also present in 2D magnetic insulators  without WM nodes, but with nonzero Berry curvature \cite{sol, sol1}. In those systems without WM nodes, the low temperature thermal Hall effect is dominant when the lowest magnon excitation is well-separated from the rest of the magnon bands. In contrast,  the  dominant contribution to the thermal Hall effect  in the current system  comes from the Berry curvature near the WM nodes at the lowest magnon excitation \cite{foot}. Moreover, also note that ferromagnetic thermal Hall effect is a consequence of the coexistence of spontaneous magnetization and out-of-plane DMI \cite{hirs, ono}. In contrast,  antiferromagnetic thermal Hall effect is neither a consequence of spontaneous magnetization nor out-of-plane DMI, because antiferromagnets have vanishingly small magnetization, and the out-of-plane DMI only stabilizes magnetic order in frustrated antiferromagnets.

\section{Conclusion}

  We have shown that stacked frustrated kagom\'e antiferromagnets are  complete topological magnon ``semimetals'', hosting both  nodal-line magnons  and triply-degenerate nodal magnon points at nonzero DMI with zero scalar spin chirality. They are transformed into  type-I and type-II Weyl magnon nodes at nonzero scalar spin chirality, and possessed a finite topological  thermal Hall effect.    The currently predicted Weyl magnon nodes do not rely on broken TRS by the magnetic order. Rather, they are provided by explicit macroscopically broken TRS by the scalar spin chirality. Therefore, the concept of Weyl nodes may exist in chiral spin liquid states \cite{her,her1}, where TRS is spontaneously broken  by the  scalar spin chirality.  As we noted above, noncoplanar chiral spin textures are intrinsic to realistic frustrated kagom\'e materials with both in-plane and out-of-plane DMI. Hence,  the present Weyl magnon nodes are, indeed, robust as all the dominant intrinsic perturbations have been taken into account. 
  
   As the Weyl magnon nodes occur at the lowest excitation they contribute immensely to the topological  thermal Hall effect as showed above. The sign of the topological  thermal Hall conductivity can be switched by a small external magnetic field, paving the way toward possible applications in magnon  spintronics and magnetic data storage devices. Moreover, a nonzero topological  thermal Hall conductivity in stacked  frustrated kagom\'e antiferromagnets could also serve as an avenue to probe  macroscopically broken time-reversal symmetry or scalar spin chirality. The predicted results can be investigated experimentally  by thermal transport measurements, and  Weyl magnon nodes can be investigated experimentally using the inelastic neutron scattering methods.
   
 We believe that our prediction of Weyl magnon nodes  in stacked (3D and quasi-2D) noncoplanar chiral spin textures is the most promising candidate towards  the first experimental realization of Weyl magnon nodes in magnetic systems.  Although we considered non-collinear spin structure with positive vector chirality,   our results should  also exist in magnetic systems with negative vector chirality (or inverse triangular spin structure) as recently reported in an insulating stacked kagom\'e antiferromagnet \cite{oku}. Similar inverse triangular spin structure was also realized in metallic frustrated magnets Mn$_3$Sn\slash Ge \cite{chen1,nak,nay}, with metallic Weyl nodes \cite{,yang, kub}, but in this case the spins cant in-plane with zero scalar spin chirality.

\textbf{Note added}. Upon arXiv submission of this manuscript, we became aware of a recent study Ref.~\cite{new},  where the authors adopted the easy-plane breathing pyrochlore antiferromagnetic model in Ref.~\cite{fei}, and studied WMs in the easy-axis counterpart with all-in-all-out (AIAO) magnetic ordering with zero scalar spin chirality. Therefore, this system also rely on broken TRS by the magnetic order and they are obviously different from the current results in noncoplanar stacked kagom\'e antiferromagnets  with nonzero scalar spin chirality.

\appendix

\section{Spin transformation}
\label{appena}
  To facilitate spin wave theory  we express the spins in terms of local axes, such that the $z$-axis coincides with the spin direction. This can be done by performing a local rotation about the $z$-axis  by the spin orientated angles $\theta_{i,\ell}$, given by 
\begin{align}
  \mathcal{R}_z(\theta_{i,\ell})=\begin{pmatrix}
\cos\theta_{i,\ell} & -\sin\theta_{i,\ell} & 0\\
\sin\theta_{i,\ell} & \cos\theta_{i,\ell} &0\\
0 & 0 &1
\end{pmatrix}.
\end{align}
Due to spin canting induced either by an external magnetic field or an in-plane DMI we perform another rotation about $y$-axis by the  angle $\vartheta$, given by
\begin{align}
  \mathcal{R}_y(\vartheta)=\begin{pmatrix}
\cos\vartheta &0& \sin\vartheta\\
0 & 1 &0\\
-\sin\vartheta &0 & \cos\vartheta
\end{pmatrix}.
\end{align}
The total rotation matrix is given by
\begin{align}
\mathcal{R}_z(\theta_{i,\ell})\cdot\mathcal{R}_y(\vartheta)
=\begin{pmatrix}
\cos\theta_{i,\ell}\cos\vartheta & -\sin\theta_{i,\ell} & \cos\theta_i\sin\vartheta\\
\sin\theta_{i,\ell}\cos\vartheta & \cos\theta_{i,\ell} &\sin\theta_{i,\ell}\sin\vartheta\\
-\sin\vartheta & 0 &\cos\vartheta
\end{pmatrix},
\end{align}
 Now, the spins transform as $ \bold{S}_i=\mathcal{R}_z(\theta_{i,\ell})\cdot\mathcal{R}_y(\vartheta)\cdot\bold S_i^\prime,$
where prime denotes the rotated frame. The classical ground state energy  is given by \begin{align}
E_{\text{cl}}&= 3NS^2\Big[ 2J\lb -\frac{1}{2} + \frac{3}{2}\cos^2\vartheta\rb -\sqrt{3}D_z\sin^2\vartheta \nonumber\\&-J_c(1-2\cos^2\vartheta)-H\cos\vartheta\Big],
\end{align}
where $N$ is the number of sites per unit cell, and the magnetic field is rescaled in unit of $S$. Minimizing this energy yields the canting angle $\cos\vartheta = H/H_s$, where $H_s=6J+2\sqrt{3}D_z+ 4J_c$ is the saturation field.

 Performing the spin transformation there are so many terms, but we will retain only the terms that contribute to noninteracting  magnon model. They are  given by

  \begin{align}
  \mathcal H_J&= J\sum_{ \la ij\ra,\ell}\big[\cos\theta_{ij,\ell} \bold{ S}_{i,\ell}^\prime\cdot \bold{ S}_{j,\ell}^\prime+ \sin\theta_{ij,\ell}\cos\vartheta \hat{\bold z}\cdot\lb \bold{ S}_{i,\ell}^\prime\times\bold{ S}_{j,\ell}^\prime\rb \nonumber\\& +2\sin^2\lb\frac{\theta_{ij,\ell}}{2}\rb\lb\sin^2\vartheta  S_{i,\ell}^{\prime x}S_{j,\ell}^{\prime x} +\cos^2\vartheta S_{i,\ell}^{\prime z} S_{j,\ell}^{\prime z}\rb\big],
 \end{align}
    
  \begin{align}   
\mathcal   H_{D_z}&= -D_z\sum_{\la ij\ra,\ell}\Big[\cos\theta_{ij,\ell}\cos\vartheta~ {\bf \hat z}\cdot \lb \bold S_{i,\ell}^\prime\times\bold S_{j,\ell}^\prime\rb\nonumber\\& - \sin\theta_{ij,\ell}\lb \cos^2\vartheta S_{i,\ell}^{\prime x}S_{j,\ell}^{\prime x} + S_{i,\ell}^{\prime y}S_{j,\ell}^{ \prime y}+\sin^2\vartheta S_{i,\ell}^{\prime z}S_{j,\ell}^{\prime z}\rb\Big],
\end{align} 

\begin{align}
 \mathcal H_{J_c}&= J_c\sum_{ i,\la \ell\ell^\prime\ra}\big[\cos\theta_{\ell\ell^\prime} \bold{ S}_{i,\ell}^\prime\cdot \bold{ S}_{i,\ell^\prime}^\prime\nonumber\\&  +2\sin^2\lb\frac{\theta_{\ell\ell^\prime}}{2}\rb\lb\sin^2\vartheta  S_{i,\ell}^{\prime x}S_{i,\ell^\prime}^{\prime x} +\cos^2\vartheta S_{i,\ell}^{\prime z} S_{i,\ell^\prime}^{\prime z}\rb\big],\\\mathcal H_Z &= -H\cos\vartheta\sum_{i,\ell} S_{i,\ell}^{\prime z},
  \end{align}
where $\theta_{\alpha\beta}=\theta_{\alpha}-\theta_{\beta}$. For antiferromagnetic interlayer coupling $J_c>0$ we have $\theta_{\ell\ell^\prime}=\pi$, whereas for ferromagnetic interlayer $J_c<0$,  $\theta_{\ell\ell^\prime}=0$. In this case only the first term in the $J_c$ term is nonzero. The scalar spin chirality of the noncoplanar  (umbrella) spin configurations is defined as \bea \chi_{ijk,l}={\bf S}_{i,\ell}^\prime\cdot( {\bf S}_{j,\ell}^\prime\times{\bf S}_{k,\ell}^\prime).\eea
Note that the scalar spin chirality is induced only within the kagom\'e  planes. 

\section{Holstein-Primakoff  transformation}
\label{appenb}
 The transformed Hamiltonian in Appendix~\eqref{appena} can now be studied by linear spin wave theory via the  Holstein-Primakoff  bosons: $
S_{i,\ell}^{z}= S-a_{i,\ell}^\dagger a_{i,\ell},~  S_{i,\ell}^{+} \approx  \sqrt{2S}a_{i,\ell}=(S_{i,\ell}^{-})^\dg$, where $ S_{i,\ell}^{\pm}=S_{i,\ell}^{x}\pm i S_{i,\ell}^{y}$ and $a_{i,\ell}^\dagger(a_{i,\ell})$ are the bosonic creation (annihilation) operators. In the following we consider the case $J_c>0$. The case $J_c<0$ can be derived in a similar way. The magnon hopping Hamiltonians are given by

\begin{align}
\mathcal H_{J-D_z}&= S\sum_{\la ij\ra,\ell}\big[ t^z(a_{i,\ell}^\dg a_{i,\ell} +a_{j,\ell}^\dg a_{j,\ell})\nonumber\\& + t^r(e^{-i\phi_{ij,\ell}}a_{i,\ell}^\dg a_{j,\ell} + h.c.)+ t^o(a_{i,\ell}^\dg a_{j,\ell}^\dg + h.c.)\big]
\end{align}

\begin{align}
\mathcal H_{J_c}&= S\sum_{i,\ell} t_c^za_{i,\ell}^\dg a_{i,\ell}\nonumber\\&  + S\sum_{i,\la \ell\ell^\prime\ra}\big[ t_c^r(a_{i,\ell}^\dg a_{i,\ell^\prime} + h.c.) + t_c^o(a_{i,\ell}^\dg a_{i,\ell^\prime}^\dg + h.c.)\big],
\end{align}

\begin{align}
  \mathcal H_{Z}&=H\cos\vartheta\sum_{i,\ell}a_{i,\ell}^\dg a_{i,\ell}
\end{align}
The  solid angle subtended by three noncoplanar spins  is given by $\phi_{ij}=\pm\phi$, where $\phi=\tan^{-1}[t_{2}^r/t_{1}^r]$. The parameters of the tight binding model are given by 
\begin{align}
& t^z= -\Big[J\lb -\frac{1}{2} + \frac{3}{2}\cos^2\vartheta\rb -\frac{\sqrt{3}}{2}D_z\sin^2\vartheta\Big],\\&
t^r=\sqrt{(t^{r}_1)^2+(t^{r}_2)^2},\\
&t_1^r=J\Big[-\frac{1}{2}+\frac{3}{4}\sin^2\vartheta\Big]-\frac{\sqrt{3}D_z}{2}\lb 1-\frac{\sin^2\vartheta}{2}\rb,\\
&t_{2}^r= -\frac{\cos\vartheta}{2}( \sqrt{3}J -D_z),\\
&t^o=\frac{\sin^2\vartheta}{4}(3J+\sqrt{3}D_z),~\\
& t_c^z= -2J_{c}\cos 2\vartheta,~
t_c^r=-  J_{c}\cos^2\vartheta,~
t_c^o=J_{c} \sin^2\vartheta.
\end{align}

\begin{figure*}
\centering
\includegraphics[width=1\linewidth]{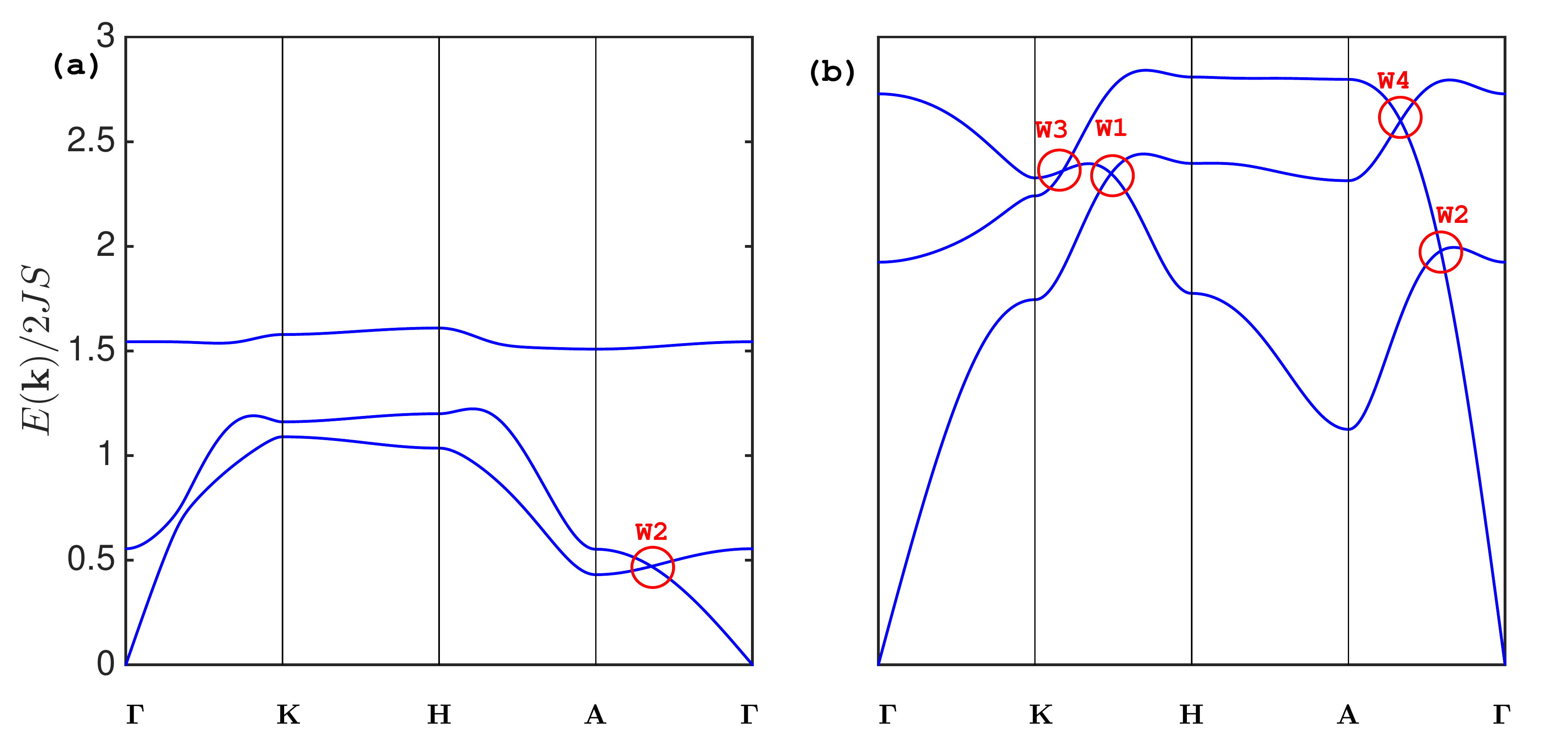}
\caption{Bulk Weyl  magnon bands of antiferromagnetically coupled stacked kagom\'e antiferromagets in the noncoplanar regime with nonzero scalar spin chirality. Here $D_z/J=0.2$,  $H=0.3H_s$.  { (a)}  $J_c/J=0.05$ (quasi-2D limit).  { (b)} $J_c/J=1.2$ (strong 3D limit). Notice that  W$_3$ is type-II WM node.}
\label{bandss}
\end{figure*}

\section{Magnon Hamiltonian}
\label{appenca}
To derive the magnon Hamiltonian we perform  Fourier transform of the Hamiltonian in Appendix~\eqref{appenb} into momentum space.  The resulting  Hamiltonian in momentum space  is given by $\mathcal H=\sum_{\bo}\psi_{k_\parallel,k_z}^\dg \mathcal{H}(k_\parallel,k_z)\psi_{k_\parallel,k_z}$, where
 
 \begin{align}
& \mathcal{H}(k_\parallel,k_z)= 2S\begin{pmatrix}
  {\boldsymbol{\mathcal{G}}^{0}}(k_z)+\boldsymbol{\mathcal{G}}^r(k_\parallel)& \boldsymbol{\mathcal{G}}^o(k_\parallel,k_z)\\
\boldsymbol{\mathcal{G}}^o(k_\parallel,k_z) &\  {\boldsymbol{\mathcal{G}}^{0}}(k_z)+\boldsymbol{\mathcal{G}}^r(k_\parallel)
\end{pmatrix}.
\label{eqnr}
\end{align}
Here, we have used the basis vector $\psi^\dg_\bo=(a_{\bo 1}^{\dg},\thinspace a_{\bo 2}^{\dg},\thinspace a_{\bo 3}^{\dg}, \thinspace a_{-\bo 1},\thinspace a_{-\bo 2},\thinspace a_{-\bo 3} )$, with $\bo=(k_\parallel,k_z)$ and $k_\parallel=(k_x,k_y)$. The  $\boldsymbol{\mathcal{G}}$ matrices are given by $\boldsymbol{\mathcal{G}}^{0}(k_z)=[\sqrt{3}D_z + J+J_c+t_c^r\cos k_z]{\bf I}_{3\times 3}$
\begin{align}
\boldsymbol{\mathcal{G}}^{r}(k_\parallel)= t^r
\begin{pmatrix}
0& \cos k_\parallel^1e^{-i\phi}& \cos k_\parallel^3 e^{i\phi}\\
\cos k_\parallel^1e^{i\phi}&0&\cos k_\parallel^2e^{-i\phi}\\
\cos k_\parallel^3e^{-i\phi}&\cos k_\parallel^2e^{i\phi}&0
\end{pmatrix},
\label{mat1}
\end{align}
\begin{align}
\boldsymbol{\mathcal{G}}^{o}(k_\parallel,k_z)= 
\begin{pmatrix}
t_c^o\cos k_z& t^o\cos k_\parallel^1& t^o\cos k_\parallel^3 \\
t^o\cos k_\parallel^1&t_c^o\cos k_z &t^o\cos k_\parallel^2\\
t^o\cos k_\parallel^3&t^o\cos k_\parallel^2&t_c^o\cos k_z
\end{pmatrix},
\label{mat2}
\end{align}
where  $k_\parallel^i=k_\parallel\cdot{\bf  a}_i$, with   ${\bf a}_1={\hat x}$, ${\bf \hat a}_2={\hat x}/2+\sqrt{3}\hat y/2$, and ${\bf \hat a}_3=-{\hat x}/2+\sqrt{3}\hat y/2$. The momentum space Hamiltonian for $J_c<0$ can be derived in a similar way.

 The magnon Hamiltonian $\mathcal{H}(k_\parallel,k_z)$  can be  diagonalized numerically using the generalized Bogoliubov transformation. This can be done  by making a linear  transformation $\psi_\bo= \mathcal{P}_\bo Q_\bo$, 
where $\mathcal{P}_\bo$ is a $2N\times 2N$ paraunitary matrix  defined as
\begin{align}
& \mathcal{P}_\bo= \begin{pmatrix}
  u_\bo& -v_\bo^* \\
-v_\bo&u_\bo^*\\  
 \end{pmatrix},
\end{align} 
where $u_\bo$ and $v_\bo$ are  $N\times N$ matrices that satisfy \bea |u_\bo|^2-|v_\bo|^2={\bf I}_{N\times N}.\eea  Here, $Q^\dg_\bo= (\mathcal{Q}_\bo^\dg,\thinspace \mathcal{Q}_{-\bo})$ with $ \mathcal{Q}_\bo^\dg=(\gamma_{\bo 1}^{\dg}, \gamma_{\bo 2}^{\dg}, \gamma_{\bo 3}^{\dg})$ being the quasiparticle operators. The matrix $\mathcal{P}_\bo$ satisfies the relations,
\begin{align}
&\mathcal{P}_\bo^\dg \mathcal{H}(\bo) \mathcal{P}_\bo=\mathcal{E}(\bo),\label{eqn1}\\ &\mathcal{P}_\bo^\dg \boldsymbol{\tau}_3 \mathcal{P}_\bo= \boldsymbol{\tau}_3,
\label{eqna}
\end{align}
where $\mathcal{E}(\bo)=\text{diag}\big[E_{n}(\bo),E_{n}(-\bo)\big]$, $\boldsymbol{\tau}_3=\text{diag}(\mathbf{I}_{N\times N}, -\mathbf{I}_{N\times N})$, and $E_{n}(\bo)$ are the  energy eigenvalues and $n$ labels the bands. Here,  `\text{diag}' denotes diagonal matrix.  From Eq.~\eqref{eqna} we get $\mathcal{P}_\bo^\dg= \boldsymbol{\tau}_3 \mathcal{P}_\bo^{-1} \boldsymbol{\tau}_3$. Therefore,  from Eq.~\eqref{eqn1} the Hamiltonian to be diagonalized  is $\mathcal{H}_B(\bo)= \boldsymbol{\tau}_3\mathcal{H}(\bo),$ whose eigenvalues are given by $ \boldsymbol{\tau}_3\mathcal E_\bo$ and the columns of $\mathcal P_\bo$ are the corresponding eigenvectors. Using the  paraunitary operator $\mathcal P_\bo$ we can define a Berry curvature as shown in the main text Eq.~\eqref{chern2}. The Chern number is the integration of Eq.~\eqref{chern2} over the BZ. 

 \begin{figure*}
\centering
\includegraphics[width=1\linewidth]{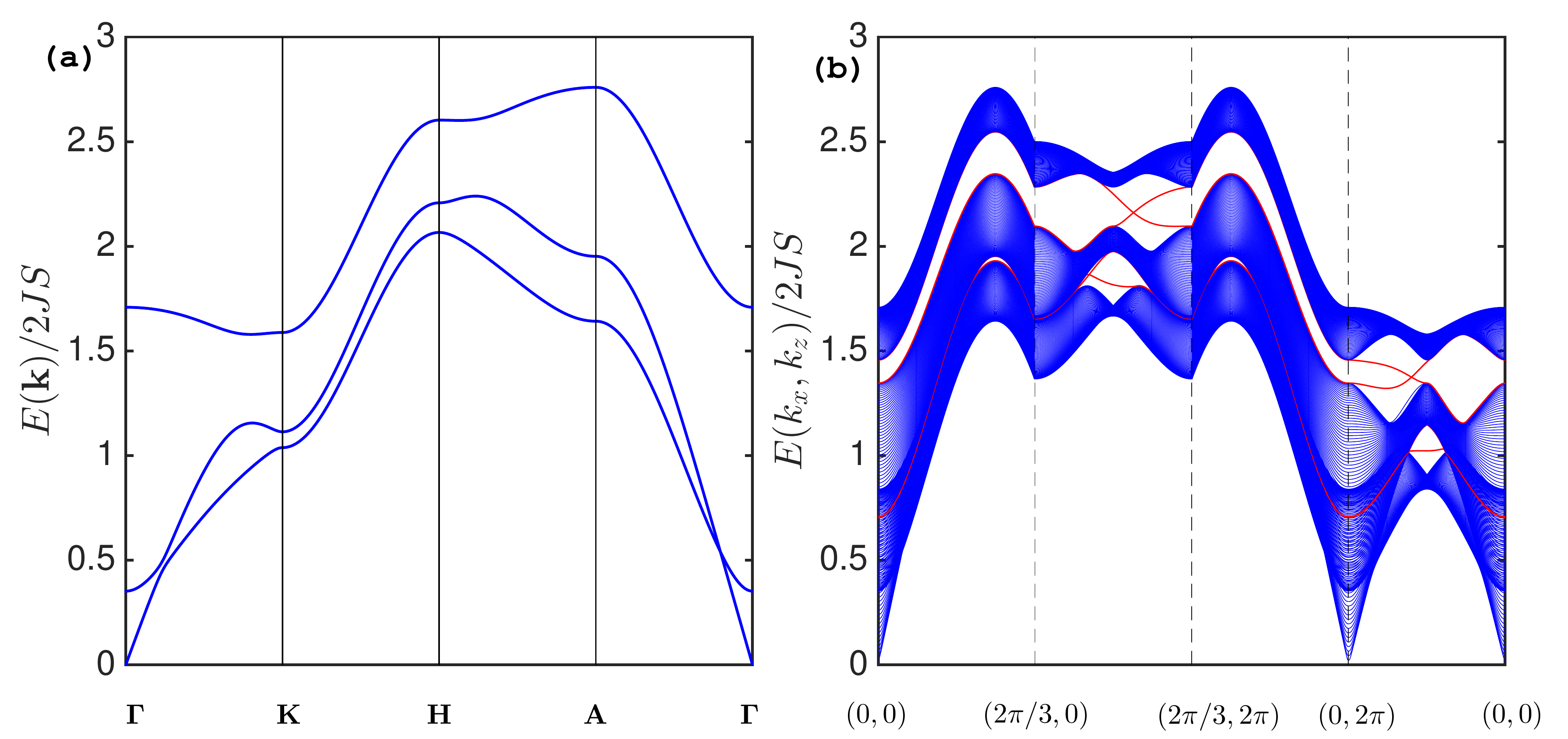}
\caption{Bulk magnon bands and surface states of ferromagnetically coupled stacked kagom\'e antiferromagets in the noncoplanar regime with nonzero scalar spin chirality. {(a)} Bulk magnon bands along the BZ paths.      {(b)} (010)-projected  magnon surface states. The parameters are  $D_z/J=0.2$, $|J_c|/J=-0.5$, $H=0.3H_s$}
\label{FM}
\end{figure*}

\section{Locations of Weyl magnon nodes}
\label{appenc}
The magnon Hamiltonian \eqref{eqnr}  cannot be diagonalized analytically, but at specific points $\Gamma_1=(k_x,k_y)=(\pm 2\pi/3,0)$ and $\Gamma_2=(k_x,k_y)=(0,0)$, the eigenvalues of the magnon Hamiltonian can be found exactly as a function of $k_z$. 

At $\Gamma_1$ the magnon energy bands are given by
\begin{align}
[E_0(k_z)]^2&=\frac{1}{2}\Big[2(\mathcal{G}^0(k_z))^2 +(t^r)^2-(t_c^o)^2-2(t^o)^2\nonumber\\&+4t_c^ot^o\cos(k_z)-(t_c^o)^2\cos(2k_z)\nonumber\\&-4t^r\mathcal{G}^0(k_z)\cos(\phi)+(t^r)^2\cos(2\phi)\Big]
\end{align}

\begin{align}
[E_{\pm}(k_z)]^2&=\frac{1}{2}\Big[(2\mathcal{G}^0(k_z))^2 +2(t^r)^2-2(t_c^o)^2-(t^o)^2\nonumber\\&-2t_c^o\lbrace 2t^o\cos(k_z)+t_c^o\cos(2k_z)\rbrace\nonumber\\&+4t^r\mathcal{G}^0(k_z)\cos(\phi)-(t^r)^2\cos(2\phi)\nonumber\\&
\pm 2\sqrt{3}t^r\sin(\phi)\lbrace 2\mathcal{G}^0(k_z) +t^r\cos(\phi)\rbrace\Big],
\end{align}
where subscript $0$ denotes lowest band, whereas $\pm$ denotes middle  and topmost bands respectively. 

At $\Gamma_2$ the magnon energy bands are given by

\begin{align}
[E_0(k_z)]^2&=\frac{1}{2}\Big[2(\mathcal{G}^0(k_z))^2 +(2t^r)^2-(t_c^o)^2-2(2t^o)^2\nonumber\\&-t_c^o\lbrace 8 t^o\cos(k_z)+t_c^r\cos(2k_z)\rbrace \nonumber\\&+8t^r\mathcal{G}^0(k_z)\cos(2\phi)+(2t^r)^2\cos(2\phi)\Big]
\end{align}

\begin{align}
[E_{\pm}(k_z)]^2&=\frac{1}{2}\Big[2(\mathcal{G}^0(k_z))^2 +(2t^r)^2-(t_c^o)^2-2(t^o)^2\nonumber\\&+4t_c^ot_c^r\cos(k_z)-(t_c^o)^2\cos(2k_z)\nonumber\\&-4t^r\mathcal{G}^0(k_z)\cos(\phi)-2(t^r)^2\cos(2\phi)\nonumber\\&
\pm 4\sqrt{3}t^r\sin(\phi)\lbrace \mathcal{G}^0(k_z) -t^r\cos(\phi)\rbrace\Big].
\end{align}
The WM nodes correspond to the points where two magnon bands cross linearly along the $k_z$ momentum direction. 

The lowest  and middle magnon  bands cross linearly at $(\pm 2\pi/3, 0, k_{W_1})$  and $(0, 0, k_{W_2})$, where $k_{W_1}=\pm \cos^{-1}(\alpha_1/\beta_1)$ and $k_{W_2}=\pm \cos^{-1}(\alpha_2/\beta_2)$, 
\begin{align}
\alpha_1 &= 3(t^o)^2 +12 t^r\cos(\phi)(1+\sqrt{3}D_z)+ 12J_ct^r\cos(\phi)\nonumber\\&-3(t^r)^2\cos(2\phi)+4\sqrt{3}t^r\sin(\phi)+12 D_zt^r\sin(\phi)\nonumber\\& +4\sqrt{3}J_ct^r\sin(\phi)+\sqrt{3}(t^r)^2\sin(2\phi),\\
\beta_1 &= 4\big[ 3t_c^ot^o-3t^rt_c^r\cos(\phi)-\sqrt{3}t^rt_c^r\sin(\phi)\big],\\
\alpha_2 &= -3(t^o)^2 + t^r\Big[ 6(1+\sqrt{3}D_z+J_c)\cos(\phi)+ 3t^r\cos(2\phi)\nonumber\\&+2 \lbrace 3D_z +\sqrt{3}(1+J_c-t^r\cos(\phi))\rbrace \sin(\phi)\Big],\\
\beta_2 &= 6t_c^ot^o-2t^rt_c^r\lbrace 3\cos(\phi)+\sqrt{3}\sin(\phi)\rbrace.
\end{align}

The topmost and lowest magnon  bands cross linearly at $(\pm 2\pi/3, 0, k_{W_3})$ and $(0,0, k_{W_4})$,  where $k_{W_3}=\pm \cos^{-1}(\alpha_3/\beta_3)$ and $k_{W_4}=\pm \cos^{-1}(\alpha_4/\beta_4)$
\begin{align}
 \alpha_3 &= 3(t^o)^2 +12 t^r\cos(\phi)(1+\sqrt{3}D_z)+ 12J_ct^r\cos(\phi)\nonumber\\&-3(t^r)^2\cos(2\phi)-4\sqrt{3}t^r\sin(\phi)-12 D_z t^r\sin(\phi)\nonumber\\& -4\sqrt{3}J_ct^r\sin(\phi)-\sqrt{3}(t^r)^2\sin(2\phi),\\
\beta_3 &= 4\big[ 3t_c^ot^o-3t^rt_c^r\cos(\phi)+\sqrt{3}t^rt_c^r\sin(\phi)\big],\\
\alpha_4 &= -3(t^o)^2 + t^r\Big[ 6(1+\sqrt{3}D_z+J_c)\cos(\phi)+ 3t^r\cos(2\phi)\nonumber\\&+2 \lbrace 3D_z -\sqrt{3}(1+J_c-t^r\cos(\phi))\rbrace \sin(\phi)\Big],\\
 \beta_4 &= 6t_c^ot^o-2t^rt_c^r\lbrace 3\cos(\phi)-\sqrt{3}\sin(\phi)\rbrace.
\end{align}

In Figs.~\ref{bandss}(a) and (b) we have shown the WM bands  in the quasi-2D and strong 3D limits respectively. We see that WM nodes persist in the noncoplanar quasi-2D limit for very small $J_c/J=0.05$ (a). However, in the strongly coupled limit (maybe unrealistic)  $J_c/J=1.2$ (b), the topmost and lowest bands form type-II WM nodes.  For ferromagnetically coupled stacked kagom\'e antiferromagnets, i.e., $J_c<0$ there are no discernible WM nodes as shown in Fig.~\eqref{FM}. Rather, the system is a topological magnon Chern insulator with gapless surface states. 



\begin{thebibliography}{99}
\bibitem{foot}
The robustness of Weyl magnons against the DMIs in pyrochlore antiferromagnets is currently not investigated.
\bibitem{foot}
   Similarly, electronic Weyl nodes close to the Fermi energy are the most important ones that make  significant contributions to the low-temperature transport properties.
\end{thebibliography}
\end{document}